\documentclass[11pt]{asaproc} 
\usepackage{graphicx}

\usepackage{times}
\usepackage{listings}

\title{A Different Approach to the Problem of Missing Data}
\author{
   Xiao (Max) Gu\thanks{Formerly a graduate student in the
   Dept. of Statistics, University of California, Davis. Now at
   Drawbridge, Inc.}
   \and
   Norman Matloff\thanks{Dept. of Computer Science, University of
   California, Davis}
}

\begin{document}

\maketitle 

\begin{abstract}

There is a long history of devleopment of methodology dealing with
missing data in statistical analysis.  Today, the most popular methods
fall into two classes, Complete Cases (CC) and Multiple Imputation (MI).
Another approach, Available Cases (AC), has occasionally been mentioned
in the research literature, in the context of linear regression
analysis, but has generally been ignored.  In this paper, we revisit
the AC method, showing that it can perform better than CC and MI, and we
extend its breadth of application.


\end{abstract}

\section{Introduction}

For concreteness in this introduction, consider a classic linear
regression analysis, based on a data matrix $D = (D_{ij})$ of $n$ rows
and $p+1$ columns, with the first $p$ columns containing the values of the
predictor variables and the last column consisting of values of the
response variable.\footnote{To simplify notation, we are assuming that
the first column consists of 1s, to accommodate a constant term in the
model}.  Some of the elements of the matrix may be missing, a condition
that is in the R language denoted by NA.

A wide variety of methods have been developed to deal with the missing
values. The most popular fall into one of two categories, again
described in our regression analysis context for convenience:

\begin{itemize}

\item {\bf Complete cases (CC):\footnote{Also known as the Listwise
Deletion method.}} Here one deletes any row in the data
matrix that has at least one NA value.

\item {\bf Multiple Imputation (MI):}  These methods involve estimating
the conditional distribution of a variable from the others, and then
sampling from that distribution via simulation. Multiple alternate
versions of the data matrix are generated, with the NA values replaced
by values that might have been the missing one.

\end{itemize}

Here we are interested in a third approach:

\begin{itemize}

\item {\bf Available Cases (AC):\footnote{Also known as the Pairwise
Deletion method.}}  If the statistical method involves computation
involving, say, various pairs of varaibles, include in such a
calculation any observation for which this pair is intact, regardless of
whether the other variables are intact.  The same holds for triples of
variables and so on.

\end{itemize}

For example, as will be detailed below, linear regression analysis only
involves computation of certain pairwise-intact values of the form

\begin{equation}
\label{dirs}
\frac{1}{n} \sum_{i=1}^n D_{ir} D_{is}
\end{equation}

Thus we may compute (\ref{dirs}) for all rows $i$ for which both
$d_{ir}$ and $d_is$ are intact --- even if some other $d_{ik}$ are
missing.  In (\ref{dirs}) the factor $1/n$ would be changed to
$1/N_{rs}$, where $N_{rs}$ is the number of rows with intact $(r,s)$
pairs in the matrix, as in for example (Cohen and Cohen, 1983).  In
other words, (\ref{dirs}) becomes

\begin{equation}
\label{newdirs}
K_{rs} =
\frac{1}{N_{rs}} 
\sum_{i=1}^n 
I_{rs} D_{ir} D_{is}
\end{equation}

where $I_{rs}$ is 1 or 0, depending on whether $D_{ir}$ and $D_{is}$
are intact.

(As noted, there are important assumptions underlying these methods, but
we defer discussion on this to Section \ref{assume}.)

Though AC was considered in the early literature on missing data, over
the years, MI methods became more and more sophisticated, and they
enjoy high popularity today. In R, for instance, there are packages {\bf
Amelia} (Honaker {\it et al}, 2011), {\bf mi} (Su {\it et al}, 2011) and
{\bf mice} (van Buuren, 2011) that apply MI techniques.  See (Little
{\it et al}, 2002) for very detailed coverage, or {\it
http://sites.stat.psu.edu/~jls/mifaq.html} for an overview.

Concurrently, interest in AC waned, not only due to its stringent
assumptions but also out of a concern that the cross products matrix
whose elements are given by (\ref{dirs}) may not be positive definite. 

We believe that AC can be a very useful tool. As Marsh notes (Marsh,
1998), AC ``follows naturally from a desire to use as much of the data
as possible.''  We will show here that AC can indeed yield significant
improvements in statistical accuracy over CC, while avoiding the very
slow computational speed of MI.  In addition to investigating the
standard AC application of linear regression nodeling, we also
investigate principal components analysis (PCA) and analysis of
contingency tables.  We make software available to implement these
methods.

\section{Choice of MI Method}

We chose {\bf Amelia} as our representative MI method, arbitrarily using
the criterion that it has the most citations on Google Scholar.  Under
the assumption that the population distribution of the rows of $D$ is
multivariate normal, an outline of its approach is as follows.  

\begin{itemize}

\item Starting the with original data $D_{ij}$, $m$ perturbations of
this data $D_{ijk}$ are created, $k = 1,...,m$, through bootstrap
sampling.  Note that these new data sets do contain NA values.

\item For $k = 1,...,m$, do:

   \begin{itemize}

   \item Replace the NAs by 0s.

   \item Use the EM algorithm and the assumption of multivariate
   normality to estimate the population mean vector and covariance
   matrix from this data.

   \item Replace each NA value by an imputed one, consisting of a value
   drawn at random from the estimated conditional distribution of this
   variable, given the intact values of the other variables.

   \end{itemize}

\item Combine the $m$ data sets, say by averaging the $m$ values of a
quantity of interest, such as a regression coefficient.  

\end{itemize}

In our initial empirical investigation, we quickly found that {\bf Amelia}
was not performing well:

\begin{itemize}

\item Its statistical accuracy was no better than those of CC and AC.

\item It was slow.  For instance, in a PCA simulation with $n = 10000$
and $p = 25$, CC and AC took 0.011 and 1.967 seconds, respectively,
while MI took 92.928 seconds.

\end{itemize}

For this reason, we will present empirical results here only for the CC
and AC methods.  It is crucial to keep in mind, though, that CC and AC
require more stringent assumptions than MI. Thus later in this paper we
will return to MI in general, and {\bf Amelia} in particular.

\section{AC in Linear Regression Models}

As noted, in the literature, AC has mostly been considered in the
context of linear regression.\footnote{Actually, to our knownledge, in
the litetature to date, AC has only been applied to covariance-related
methods, including linear regression.} Thus we will begin there.

\subsection{Motivation and Method}

Consider the case of random-X regression. Define the matrix $U$ and the
vector $V$ to be $D$ minus the last column, and the last column of $D$,
respectively.  Then the classic formula for the vector of estimated
regression coefficients, assuming intact data, is

\begin{equation}
\label{uprimeuetc}
(U'U)^{-1}(U'V) = 
\left (\frac{1}{n} U'U \right )^{-1} \left (\frac{1}{n} U'V \right ) 
\end{equation}

which as $n \rightarrow \infty$ converges to

\begin{equation}
\label{classformula}
[E(X X')]^{-1} E(XY)
\end{equation}

where the random column vector $X$ and and the random scalar variable
$Y$ have the population distribution from which the rows of $U$ and
elements of $V$ are sampled.  

The point is that this convergence still holds if in (\ref{uprimeuetc}),
we replace the $(r,s)$ element of $U'U$ in (\ref{uprimeuetc}) by
(\ref{newdirs}), $r,s = 1,...,p$ and replace element $r$ in $U'V$ by
(\ref{newdirs}) with $s = p + 1$.

\subsection{Implementation}

R code for use of AC as a replacement for {\bf lm()} is available in two
implementations (not just two locations), a function {\bf lmmv()} at
{\it https://github.com/maxguxiao/Available-Cases}, and a function {\bf
lmac()} in the {\bf regtools} package  at {\it
https://github.com/matloff/regtools}.

The latter takes advantage of the fact that R's {\bf cov()} function
offers an argument option {\bf use=pairwise.complete.obs}, which applies
AC to finding covariance matrices, which in turn can be used to estimate
regression coefficients::

\begin{verbatim}
# arguments:

# x: predictor values (no 1s column)
# y: response variable values

lmac <- function(x,y) {
   p <- ncol(x)
   tmp <- cov(cbind(x,y),use='pairwise.complete.obs')
   upu <- tmp[1:p,1:p]
   upv <- tmp[1:p,p+1]
   bhat <- solve(upu,upv)
   bhat0 <- 
      mean(y,na.rm=TRUE) - colMeans(x,na.rm=TRUE) %*% bhat
   c(bhat0,bhat)
}
\end{verbatim}

This works because for centered data, (\ref{classformula}) is equal to

\begin{equation}
Cov(X)^{-1} Cov(X,Y)
\end{equation}

Since the {\bf use=pairwise.complete.obs} option in R's {\bf cov()} uses
the AC method, this gives us AC estimation for linear regression.

The above code, {\bf lmac()}, is much faster than {\bf lmmv()}, since R's
{\bf cov()} function operates at C-level, as opposed to the use of R
{\bf for()} loops in {\bf lmmv()}.  

\subsection{Standard Errors for the Coefficients}

Our code computes standard errors for the estimated regression
coefficients in two different ways.

{\bf lmmv():}

The {\bf lmmv()} function uses the delta method, together with numerical
calculation of derivatives using the {\bf numDeriv} package.  Any
component of (\ref{uprimeuetc}) is a function of the $K_{rs}$ in
(\ref{newdirs}) for $1 \leq r \leq s \leq p+1$.  The function {\bf
genD()} in {\bf numDeriv} is then used to compute the numerical gradient
$G$ of this function.  

The standard error is then

\begin{equation}
\sqrt{G'BG}
\end{equation}

where $B$ is the estimated covariance matrix for the $K_{rs}$
(conditional on the $N_{rs}$).  We have

\begin{eqnarray}
Cov(K_{ab},K_{cd})
&=& 
\frac{1}{N_{ab}}
\frac{1}{N_{cd}}
\sum_{i=1}^n 
Cov(1_{ab}D_{ia} D_{ib},
    1_{cd}D_{ic} D_{id}) 
\end{eqnarray}

There are various ways to evaluate this, 
such as calling {\bf Cov()} with the {\bf pairwise.complete.obs} option.
% the simplest of which would be
% to apply R's {\bf cov()} to the CC data.

{\bf lmac():}

In {\bf lmac()}, we simply use the bootstrap to generate standard
errors.  Though this may seem more time-consuming than using the delta
method, this consideration is countered by the fact that {\bf genD()} is
written in R rather than C, and thus involves slow loop computation.

\subsection{Empirical Evaluation}
\label{lmemp}

We present here simulations that run on real or simulated data.  The
idea is that, starting with a given data set, in each repetition of the
simulation, random NA values are inserted, and the value of
$\widehat{\beta}_1$ is recorded. The variance of such values for {\bf
lmac()} is compared to that for {\bf lm()}; the latter represents CC, as
its method of handling NAs is CC.\footnote{The means values for the two
methods were virtually identical.}

We tried it for several real data sets.  One is the Pima study 
at the UCI Machine Learning Repository,
{\it https://archive.ics.uci.edu/ml/datasets/Pima+Indians+Diabetes}.
Here $n = 768$ and $p = 8$.  We took blood pressure as our response
variable, and all the other variables as predictors.  Results for
inserting 1\%, 5\% and 10\% NAs were as shown in Table \ref{pima}.

\begin{table}
\begin{center}
\vskip 0.5in
\begin{tabular}{|r|r|r|}
\hline
NA rate & CC var. & AC var. \\ \hline 
0.01 & 0.008034006 & 0.002094305 \\ \hline 
0.05 & 0.05018815 & 0.01230746 \\ \hline
0.10 & 0.1421812 & 0.02398466 \\ \hline
\end{tabular}
\end{center}
\caption{Pima Data, Linear Regression}
\label{pima}
\end{table}

AC was much more accurate, especially with the heavier NA rate.

We also tried the method on some Census data, concerning programmers and
engineers in Silicon Valley.  (This data set is available in the {\bf
regtools} package.)  Here, $n = 20090$ and $p = 11$.  The results are
shown in Table \ref{prgeng}.

\begin{table}
\begin{center}
\vskip 0.5in
\begin{tabular}{|r|r|r|}
\hline
NA rate & CC var. & AC var. \\ \hline 
0.01 & 0.4694873 & 0.1387395 \\ \hline
0.05 & 2.998764 & 0.7655222 \\ \hline
0.10 & 8.821311 & 1.530692 \\ \hline
\end{tabular}
\end{center}
\caption{Census Data, Linear Regression}
\label{prgeng}
\end{table}

Next, we considered the baseball player data set in CRAN's {\bf
freqparcoord} package (Matloff and Xie, 2014), which consists of data on
height, weight, age and position for 1015 major league
players.\footnote{Data courtesy of the UCLA Statistics Department.}  In
predicting weight from only height and age, there appeared to be no real
difference in the accuracy of CC and AC; see Table \ref{mlb1}.  However,
when playing position was added to the prediction, with dummy variables
for infielders, outfielders and pitchers,\footnote{The remaining
categories are catchers and, in the American League, designated
hitters.} AC greatly outperformed CC, as seen in Table \ref{mlb2}.

\begin{table}
\begin{center}
\vskip 0.5in
\begin{tabular}{|r|r|r|}
\hline
NA rate & CC var. & AC var. \\ \hline 
0.01 & 0.001587028 & 0.001587711 \\ \hline
0.05 & 0.009455012 & 0.009799962 \\ \hline
0.10 & 0.02019519 & 0.01996154 \\ \hline
\end{tabular}
\end{center}
\caption{Baseball Data I, Linear Regression}
\label{mlb1}
\end{table}

\begin{table}
\begin{center}
\vskip 0.5in
\begin{tabular}{|r|r|r|}
\hline
NA rate & CC var. & AC var. \\ \hline 
0.01 & 0.00354029 & 0.00211546 \\ \hline
0.05 & 0.02160327 & 0.01146491 \\ \hline
0.10 & 0.05171839 & 0.02553519 \\ \hline
\end{tabular}
\end{center}
\caption{Baseball Data II, Linear Regression}
\label{mlb2}
\end{table}

In all cases, AC did quite well. However, we also
compared CC and AC on data generated as

\begin{verbatim}
n <- 2500
p <- 2
p1 <- p + 1 
a <- 5
b <- 8
ones <- matrix(rep(1,p),ncol=1)
z <- matrix(nrow = n, ncol = p1)
z[,1:p] <- runif(n*p,min=a,max=b)
z[,p1] <- 
   z[,1:p] %*% ones + sgm * runif(n,min = -0.5,max = 0.5)
\end{verbatim}

As seen in Table \ref{simlm} AC does eventually dominate, but only for
the larger value of {\bf sgm}, and AC does considerably worse than CC
before that.

\begin{table}
\begin{center}
\vskip 0.5in
\begin{tabular}{|r|r|r|r|}
\hline
NA rate & CC var. & AC var. & sgm \\ \hline 
0.01 & 5.381073e-06 & 8.068427e-05 & 1 \\ \hline
0.10 & 0.0001015785 & 0.0009777008 & 1 \\ \hline
0.10 & 0.002390376 & 0.001291069 & 5 \\ \hline
\end{tabular}
\end{center}
\caption{Simulated Data, PCA}
\label{simlm}
\end{table}

\section{AC in Principal Components Analysis}

Once one uses AC in the context of covariance matrices for linear
regression analysis, it is natural to do so for PCA.  The {\bf regtools}
version, {\bf pcac()}, is quite simple:

\begin{verbatim}
pcac <- function (indata, scale = FALSE) 
{
    covcor <- if (scale) 
        cor
    else cov
    cvr <- covcor(indata, use = "pairwise.complete.obs")
    tmp <- eigen(cvr)
    res <- list()
    if (any(tmp$values < 0)) 
        stop("at least one negative eigenvalue")
    res$sdev <- sqrt(tmp$values)
    res$rotation <- tmp$vectors
    res
}
\end{verbatim}

The quantity of interest was the square root of the maximal eigenvalue.
AC was much more effective than CC on the
Pima (Table \ref{pimapc}),
Census (Table \ref{pepc}) and
baseball (Table \ref{mlbpc}) data.

\begin{table}
\begin{center}
\vskip 0.5in
\begin{tabular}{|r|r|r|}
\hline
NA rate & CC var. & AC var. \\ \hline 
0.01 & 3.860661 & 0.3721266 \\ \hline
0.05 & 23.8738 & 1.976418 \\ \hline
0.10 & 64.26592 & 4.95431 \\ \hline
\end{tabular}
\end{center}
\caption{Pima Data, PCA}
\label{pimapc}
\end{table}

\begin{table}
\begin{center}
\vskip 0.5in
\begin{tabular}{|r|r|r|}
\hline
NA rate & CC var. & AC var. \\ \hline 
0.01 & 32403.34 & 4498.546 \\ \hline
0.05 & 147780.2 & 20018.99 \\ \hline
0.10 & 562266.5 & 64522.77 \\ \hline
\end{tabular}
\end{center}
\caption{Census Data, PCA}
\label{pepc}
\end{table}

\begin{table}
\begin{center}
\vskip 0.5in
\begin{tabular}{|r|r|r|}
\hline
NA rate & CC var. & AC var. \\ \hline 
0.01 & 0.01391677 & 0.002439572 \\ \hline
0.05 & 0.07892307 & 0.01110466 \\ \hline
0.10 & 0.2025108 & 0.02432591 \\ \hline
\end{tabular}
\end{center}
\caption{Baseball Data, PCA}
\label{mlbpc}
\end{table}

\begin{table}
\begin{center}
\vskip 0.5in
\begin{tabular}{|r|r|r|}
\hline
NA rate & CC var. & AC var. \\ \hline 
0.01 & 0.0001565521 & 1.412733e-05 \\ \hline
0.05 & 0.001146238 & 7.169807e-05 \\ \hline
0.10 & 0.004132952 & 0.0001567328 \\ \hline
\end{tabular}
\end{center}
\caption{Simulated Data, PCA}
\label{simpc}
\end{table}

For the simulated data {\bf z}, as in Section \ref{lmemp}, the results
were a little different, with AC still doing very well, but with a
caveat. 
AC performed well, as seen in Table \ref{simpc}.  But it sometimes
failed, due to negative eigenvalues.  Of 100 trials for each of the NA
rates of 0.01, 0.05 and 0.10, there were 0, 7 and 13 instances of
negative eigenvalues.  This is related to the concern about possible
lack of positive definiteness mentioned earlier. 

It must be noted that this distribution is highly artificial. The square
root of the population maximal eigenvalue is about 2.88, quite small in
comparison to the population mean of about 65 for the last variable in
the data.  Nevertheless, the above results should be kept in mind.

\section{AC in the Log-Linear Model}

To our knowledge, this is the first attempt to use AC outside of the
realm of estimation of covariance matrices.\footnote{By contrast, there
is an extenisve litetature on MI methods for generalized linear models,
including the log-linear model. See (Ibrahim {\it et al}, 2005). Note by
the way that {\bf Amelia} is not appropriate for this setting, due to
its assumptuon of multivariate normality for the data.}

\subsection{Motivation and Method}

We will illustrate the method here in the 3-factor setting, using the
formulations of (Christensen, 1998, Chapter 3).  Call the factorx $X$,
$Y$ and $Z$.

As our example computation, take the model in which $X$ and $Y$ are
conditionally independent, given $Z$.  Then the probability of an
individual falling into cell $ijk$ is

\begin{eqnarray}
p_{ijk} 
&=& P(X = i, Y = j, Z = k) \\ 
&=& P(Z = k) ~ P(X = i, Y = j ~|~ Z = k) \\
&=& P(Z = k) ~ P(X = i ~|~ Z = k) ~ P(Y = j ~|~ Z = k) \\
&=& 
\frac
{p_{i.k}
 ~ p_{.jk}} 
{p_{..k}}
\end{eqnarray}

This is a perfect opportunity for AC. For instance, we estimate $p_{i.k}$
as

\begin{equation}
\widehat{p}_{i.k} =
\frac{1}{N_{i.k}} \sum_{m=1}^n 1_{X_m = i, Z_m = k}
\end{equation}

where $N_{i.k}$ is the number of data points in which $X$ and $Z$ are
intact.

\subsection{Implementation}

This is all implemented in the function {\bf loglinac()} in {\bf
regtools}.\footnote{The log-linear model portion of {\bf regtools} is
just a prototype. At present, it handles only the 3-factor case, and
does only point estimation.}  It works as follows.

First, AC is used to estimate all the model quantities, e.g.  $p_{i.k}$
above. These are all multipled by the total number of observations,
yielding estimated expected cell frequencies. The latter are then
treated as ``observed cell counts,'' and fed into R's {\bf loglin()}
function. These produce the correct log-linear model
coefficients.\footnote{Of course, numbers such as the Pearson's test
that come out of this are not valid.}

Though {\bf loglin()} takes its input data in the form of an R table, in
order to use AC we need a data frame. The function {\bf tbltofakedf()}
creates a dataframe from a table for this purpose.

\subsection{Empirical Evaluation}

Here we used the {\bf UCBAdmissions} table built-in to R.  Continuing
with the above example, the model used was that in which the factors
Admitted and Gender were conditionally independent, given Department.
The call is

\begin{verbatim}
uca <- tbltofakedf(UCBAdmissions)
loglinac(uca,list(c(1,3),c(2,3)))
\end{verbatim}

As seen in Table \ref{ucb}, once again AC can yield substantial
improvements.

\begin{table}
\begin{center}
\vskip 0.5in
\begin{tabular}{|r|r|r|}
\hline
NA rate & CC var. & AC var. \\ \hline 
0.01 & 4.395758e-05 & 2.781903e-05 \\ \hline
0.05 & 0.0002362016 & 0.0001513719 \\ \hline
0.10 & 0.0005367046 & 0.000360953 \\ \hline
\end{tabular}
\end{center}
\caption{UCB Admissions Data, Log-Linear Model}
\label{ucb}
\end{table}

\section{Back to the MI Issue}
\label{assume}

As mentioned, there has been concern about AC in two senses:  positive
definiteness of covariance matrices, and stringency of assumptions. Here
we revisit both of these issues.

\subsection{Positive Definite Covariance Matrices}

One of the concerns that have arisen for the AC method in the past was
possible lack of positive definiteness of $U'U$ in (\ref{uprimeuetc}).
Yet Marsh found that this is rarely a problem (Marsh, 1998).\footnote{As
noted earlier, though, we did occasionally encounter negative
eigenvalues in the simulated data in our PCA studdy.} 

Furthermore, the problem can occur in {\bf Amelia} as well.  This is
because the procedure replaces NA values in the bootstrapped versions of
the original data by 0s.  Indeed, the {\bf Amelia} code does include
checks for this, halting the procedure upon detection of a problem.

\subsection{Assumptions}

The issue of assumptions is more delicate, especially since, as is well
recognized, the assumptions involved with CC, AC and MI are difficult
to check using the data.

Let $Y$ denote a variable of interest, and let $M$ be 1 or 0, depending
on whether $Y$ is missing.  Also, let $D$ denote the vector of the
other varialbes, which for simplicity we assume are never missing.  For
the same reason, we also assume the variables are discrete-valued rather
than continuous.

CC and AC assume a Missing Completely at Random (MCAR) setting, which is
usually defined as something like\footnote{There is some variation in
the literature on the details of the assumptions discussed here.}

\begin{equation}
\label{mcar}
P(M = 1 | Y = s, D = t) = P(M = 1)
\end{equation}

where $t$ is in general vector-valued.  Thus $M$ is independent of
$(Y,D)$.  Turning this around, we have

\begin{equation}
P(Y = s, D = t | M = i) =
P(Y = s, D = t) 
\end{equation}

for $i = 0, 1$.  In other words, the distribution of $(Y,D)$ is the same,
whether $Y$ is missing or not, and thus inference made from the cases in
which $Y$ is observed generalize properly to the full distribution of
$(Y,D)$.

MI assumes somewhat less, a condition known as Missing at Random (MAR).
In our context here, this is defined as

\begin{equation}
\label{mar}
P(M = 1 | Y = s, D = t) = P(M = 1 | D = t)
\end{equation}

A typical example of the idea behind MAR is given in (Cohen and Cohen,
1983), concerning a study of student motivation in a classroom survey.
We might surmise that students who have low levels of motivation are
less likely to answer the survey question, $Y$, concerning their level
of motivation.  But other factors $D$, such as socioeconomic status may
explain $Y$ so well that (\ref{mar}) holds. The problem of course is
that the predictive ability of $D$ may not be strong enough to justify
(\ref{mar}).  Moreover, in practice some of the values in the vector $D$
will also be missing, further weakening the MAR assumption. 

Also, in the case of {\bf Amelia} in particular, recall that in its EM
computations, it replaces NA values by 0s, possibly producing further
bias.

The literature on missing data often includes casual comments to the
effect that use of CC in settings in which MCAR fails, but in which MAR
holds, results in bias.  Actually, this is not necessarily the case, as
will be discussed in the next two sections.  Though some careful
treatments exist for the regression case, such as (Glynn and Laird,
1986), the analysis here will go into greater generality, i.e.\ will not
be limited to expected values, and in any case is simple enough to
include here.

\subsubsection{Estimation of Conditional Quantities Under MAR}

Let's see what happens under MAR in the case of regression analyses and
other types of association analysis.

Rewrite (\ref{mar}) as

\begin{eqnarray}
P(Y = s | D = t, M = i) ) 
&=& 
\frac
{P(Y = s,  D = t, M = i)}
{P(D =t, M = i)}
\nonumber
\\
&=& 
\frac
{P(M = i | Y = s,  D = t) ~
 P(Y = s,  D = t)}
{P(D = t, M = i)} 
\nonumber
\\
&=& 
\frac
{P(M = i | Y = s,  D = t) ~
 P(Y = s | D = t) ~ P(D = t)}
{P(D = t, M = i)} 
\nonumber
\\
&=& 
\frac
{P(M = i | D = t) ~
 P(Y = s | D = t) ~ P(D = t)}
{P(D = t, M = i)} 
\\
&=& P(Y = s | D = t) 
\label{sameconddist}
\end{eqnarray}

where the next-to-last equality comes from (\ref{mar}).

In other words, if we are interested in the relation between $Y$ and
$D$, say by performing regression analysis of $Y$ on $D$ --- i.e.\
modeling the conditional distribution of $Y$ given $D$ --- our being
deprived of the missing values of $Y$ will not bias our regression
analysis.

In fact, (\ref{sameconddist}) has the rather ironic implication:

\begin{quote}
The MAR assumption is meant to apply to situations in which CC and AC
ostensibly cannot be used.  Yet, if our goal is regression analysis or
other types of measures of association, CC and AC can indeed be used
in MAR settings after all.
\end{quote}

\subsubsection{Estimation of Unconditional Quantities Under MAR}

On the other hand,

\begin{eqnarray}
P(Y = s | M = 0) &=&
\frac{P(Y = s, M = 0)}{P(M = 0)} \\
&=&
\label{diffunconddist}
\frac{P(M = 0|Y = s)}{P(M = 0)} \cdot P(Y = s)
\end{eqnarray}

In other words, our estimate of $P(Y = s)$, an unconditional quantity,
may be biased upward or downward.  Take the student motivation example,
for instance.  For values of $s$ coding high motivation, we surmise in
(\ref{diffunconddist}),

\begin{equation}
\frac{P(M = 0|Y = s)}{P(M = 0)} > 1
\end{equation}

thus causing an upward bias in the intact data.

\section{Conclusions and Future Work}

This work has found the following:

\begin{itemize}

\item Studies on various real data sets were presented here that showed
that (under the MCAR assumption), AC can greatly outperform CC.

\item Although MI is thought of as a method to use when AC's MCAR
assumption does not hold, under MI's MAR assumption, AC still produces
statistically correct results for regression analyses and other models
of association.

\item Situations in which MAR holds but MCAR does not may be rather
rare.

\item MI computation is extremely slow, and does not seem to be any
better statistically than AC. 

\item Thus, for regression/association analysis, AC may actually be a
competitive alternative to MI.

\end{itemize}

Clearly, though, these conclusions are tentative.  Much more
investigation needs to be done on MI, including its statistical
efficiency relative to AC.

\end{document}